\documentstyle[12pt]{article}

\renewcommand{\theequation}{\thesection.\arabic{equation}}

\newlength{\extraspace}
\newlength{\extraspaces}
\newcounter{dummy}

\newcommand{\baa}{
\addtocounter{equation}{1}
\setcounter{dummy}{\value{equation}}
\setcounter{equation}{0}
\renewcommand{\theequation}{\thesection.\arabic{dummy}\alph{equation}}
\begin{eqnarray}
\addtolength{\abovedisplayskip}{\extraspaces}
\addtolength{\belowdisplayskip}{\extraspaces}
\addtolength{\abovedisplayshortskip}{\extraspace}
\addtolength{\belowdisplayshortskip}{\extraspace}}

\newcommand{\eaa}{
\end{eqnarray}
\setcounter{equation}{\value{dummy}}
\renewcommand{\theequation}{\thesection.\arabic{equation}}}

\newcommand{\be}{\begin{equation}
\addtolength{\abovedisplayskip}{\extraspaces}
\addtolength{\belowdisplayskip}{\extraspaces}
\addtolength{\abovedisplayshortskip}{\extraspace}
\addtolength{\belowdisplayshortskip}{\extraspace}}
\newcommand{\ee}{\end{equation}}

\newcommand{\ba}{\begin{eqnarray}
\addtolength{\abovedisplayskip}{\extraspaces}
\addtolength{\belowdisplayskip}{\extraspaces}
\addtolength{\abovedisplayshortskip}{\extraspace}
\addtolength{\belowdisplayshortskip}{\extraspace}}
\newcommand{\ea}{\end{eqnarray}}

\newcommand{\bd}{\begin{displaymath}
\addtolength{\abovedisplayskip}{\extraspaces}
\addtolength{\belowdisplayskip}{\extraspaces}
\addtolength{\abovedisplayshortskip}{\extraspace}
\addtolength{\belowdisplayshortskip}{\extraspace}}
\newcommand{\ed}{\end{displaymath}}

\newcommand{\ban}{\begin{eqnarray*}
\addtolength{\abovedisplayskip}{\extraspaces}
\addtolength{\belowdisplayskip}{\extraspaces}
\addtolength{\abovedisplayshortskip}{\extraspace}
\addtolength{\belowdisplayshortskip}{\extraspace}}
\newcommand{\ean}{\end{eqnarray*}}

\newcommand{\newsection}[1]{
\vspace{15mm}
\pagebreak[3]
\addtocounter{section}{1}
\setcounter{equation}{0}
\setcounter{subsection}{0}
\setcounter{footnote}{0}
\begin{center}
{\Large \thesection. #1}
\end{center}
\nopagebreak
\medskip
\nopagebreak}

\newcommand{\nonu}{\nonumber \\[.5mm]}

\newcommand{\deel}[2]{{\textstyle{#1 \over #2}}}
\newcommand{\hf}{{\textstyle{1\over 2}}}

\newcommand{\ie}{{\it i.e.}}

\newcommand{\re}{\mbox{I}\!\mbox{R}}

\def\inbar{\,\vrule height1.5ex width.4pt depth0pt}
\font\rms=cmr12 at 12pt
\def\ce{\relax\ifmmode\mathchoice
{\hbox{$\inbar\kern-.3em{\rm C}$}}
{\hbox{$\inbar\kern-.3em{\rm C}$}}
{\lower.9pt\hbox{\rms $\inbar\kern-.3em{\rm C}$}}
{\lower1.2pt\hbox{\rms $\inbar\kern-.3em{\rm C}$}}
\else{$\inbar\kern-.3em{\rm C}$}\fi}
\font\cmss=cmss12 \font\cmsss=cmss12 at 12pt
\def\ze{\relax\ifmmode\mathchoice
{\hbox{\cmss Z\kern-.4em Z}}{\hbox{\cmss Z\kern-.4em Z}}
{\lower.9pt\hbox{\cmsss Z\kern-.4em Z}}
{\lower1.2pt\hbox{\cmsss Z\kern-.4em Z}}\else{\cmss Z\kern-.4em Z}\fi}

\newcommand{\dif}{\partial}
\newcommand{\dbar}{\bar{\dif}}

\newcommand{\abar}{A_{\bar{z}}}

\newcommand{\gam}{\Gamma}

\newcommand{\tr}{\mbox{Tr}}
\newcommand{\del}{\delta}

\newcommand{\accc}[1]{\deel{1}{\pi}\int d^2z \, \tr ( {#1} ) }
\newcommand{\actie}[1]{\deel{1}{2\pi}\int d^2z \, }
\newcommand{\aksie}{\deel{1}{\pi}\int d^2z \, }
\newcommand{\aksies}[1]{\deel{#1}{\pi}\int d^2z \, }

\newcommand{\vars}[2]{{\del {#1} \over \del {#2}}}

\newcommand{\mat}[9]{\left( \begin{array}{ccc}
				#1 & #2 & #3 \\
				#4 & #5 & #6 \\
				#7 & #8 & #9
                             \end{array} \right) }

\newcommand{\np}[1]{Nucl. Phys. {\bf B#1}}
\newcommand{\cmp}[1]{Comm. Math. Phys. {\bf #1}}

\newcommand{\plb}[1]{Phys. Lett. {\bf B#1}}

\newcommand{\add}{\mbox{\rm ad}}

\newcommand{\eps}{\epsilon}

\begin{document}
\addtolength{\baselineskip}{.7mm}

\thispagestyle{empty}
\begin{flushright}
{\sc THU}-92/33\\
11/92
\end{flushright}
\vspace{1.5cm}
\setcounter{footnote}{2}
\begin{center}
{\LARGE\sc{The Effective Action of $W_3$ Gravity to All Orders}}\\[2cm]

\sc{Jan de Boer\footnote{e-mail: deboer@ruunts.fys.ruu.nl}
and Jacob Goeree\footnote{e-mail: goeree@ruunts.fys.ruu.nl}}
\\[8mm]
{\it Institute for Theoretical Physics\\[2mm]
University of Utrecht\\[2mm]
Princetonplein 5\\[2mm]
P.O. Box 80.006\\[2mm]
3508 TA Utrecht}\\[1.5cm]

{\sc Abstract}\\[1cm]
\end{center}

\noindent
The effective action for chiral $W_3$ gravity is studied. It is shown
that the computation of the effective action can be reduced to
that of a $SL(3,\re)$
Wess-Zumino-Witten theory. If one assumes that the effective
action for the Wess-Zumino-Witten model is identical to the
WZW action up to multiplicative renormalizations, then the
effective action for $W_3$ gravity is, to all orders, given by a
constrained WZW model. The multiplicative renormalization
constants of the WZW model are discussed and it is analyzed
which particular values of these
constants are consistent with previous
one-loop calculations, and which reproduce the KPZ formulas for
gravity and their generalizations for $W_3$ gravity.

\vfill

\newpage

\newsection{Introduction}

Starting with a two-dimensional conformal field theory
coupled to gravity one can construct
an induced action for pure gravity by integrating out
the matter degrees of freedom from the theory.
One of the striking features of two dimensions is that
the form of the induced action does not
depend on the detailed form of the field theory, but only on its
central charge. Therefore, this induced action is a good
starting point to study the general properties of
two-dimensional quantum gravity.
The quantization of
the induced action can proceed in several ways, depending on the
gauge condition that one imposes on the
remaining symmetries of the induced gravity action.
In the conformal gauge this leads to
Liouville theory, whereas in the chiral gauge the resulting
theory is non-local.
In this paper we deal with the quantization
of the chiral induced action of $W$ gravity. We do not
discuss the quantization of the covariant
induced action for $W$ gravity
in the conformal gauge which, as was demonstrated in \cite{jj,jj2},
would amount to the quantization of Toda theory. More on $W$
gravity can be found in \cite{wreview1,wreview2,wreview3,wreview4}
and references therein.

For ordinary gravity, the chiral induced action can be defined
by
\be \label{def:sindgr}
e^{-\Gamma_c[\mu]}=\left\langle e^{-\aksie\mu T }
\right\rangle_{OPE},
\ee
where the right hand side is computed using the operator product
expansion of $T$ with itself. Here, the only remnant of the
field theory one started with is the central charge $c$
occurring in
the operator product expansion; $\mu$ refers to a
component of the metric in the chiral gauge, $ds^2=dz d\bar{z} +
\mu d\bar{z} d\bar{z}$, and for this reason $\mu$ is sometimes
denoted by $h_{\bar{z}\bar{z}}$.
To quantize this chiral induced
action, we want to compute the generating functional for
correlation functions of $\mu$,
\be \label{def:seffgr}
e^{-\Gamma_{eff}[T]}=\int {\cal D} \mu
 e^{\aksie\mu T-\Gamma_c[\mu]}.
\ee
This functional $\Gamma_{eff}[T]$ can be expanded in terms of
$1/c$, $\Gamma_{eff}[T]=\sum_{i\geq 0} c^{1-i}
\Gamma_{eff}^{(i)}[T]$, and $\Gamma_{eff}^{(0)}$ is the Legendre
transform of $\Gamma_c[\mu]$. As the precise form of
$\Gamma_c[\mu]$ is known \cite{pol2},
\be \label{exactgr}
\Gamma_c[\mu]=-\frac{c}{24\pi}\int d^2 z \, \mu\,\dif^2\,
(1-\frac{1}{\dbar}\,\mu\,\dif)^{-1}\, \frac{\dif}{\dbar}\, \mu,
\ee
one can in principle
compute $\Gamma_{eff}[T]$ order by order. This has been done up
to one loop \cite{meipav,zam,kpz,grini},
and the one loop result is
$-25\Gamma_{eff}^{(0)}+13 T\vars{\Gamma_{eff}^{(0)}}{T}$,
so that the total result up to one loop can be written as
$(c-25)\Gamma_{eff}^{(0)}[(1+\frac{13}{c})T]$. This
suggests \cite{zam}
that the all order result for $\Gamma_{eff}[T]$ is given
by $Z_k c \Gamma_{eff}^{(0)}[Z_T T]$, for certain constants
$Z_k$ and $Z_c$ that are power series in $1/c$. However, it is
clear that it will become more and more cumbersome to go to
higher orders, so we will choose a different strategy to compute
$\Gamma_{eff}[T]$.

Instead of using (\ref{def:sindgr}) as a starting point, we
could also start with the following definition of
$\Gamma_c[\mu]$,
\be \label{def:sindgr2}
e^{-\Gamma_c[\mu]}=\int {\cal D}\phi e^{-S(\phi) -\aksie\mu
T(\phi)},
\ee
where $T(\phi)$ is such that upon quantizing the $\phi$ degrees
of freedom, $T(\phi)$ becomes an energy momentum tensor with
central charge $c$. One now immediately finds that
\be \label{def:seffgr2}
e^{-\Gamma_{eff}[T]}=\int {\cal D}\phi e^{-S(\phi)}
\delta(T-T(\phi)).
\ee
In general, it is very difficult to perform the path integral
over the $\phi$ fields in the presence of this delta function,
but it turns out that if we start with a constrained $Sl(2,\re)$
WZW theory as an action, it is possible to perform this path
integral and thus compute the effective action for gravity to
all orders. This construction is closely related to the 'hidden
$SL(2,\re)$ symmetry' in 2-d gravity \cite{beroog,pol2,kpz}. In
the constrained $SL(2,\re)$ WZW model one of the currents
becomes, upon imposing the constraints, the energy momentum
tensor of the theory, and therefore the delta function in
(\ref{def:seffgr2}) becomes a delta function for a current of
the WZW theory and can be integrated out. For this last step one
has to perform a change of variables in the WZW theory from the
group variable $g$ to the current $g^{-1}\dif g$, and take the
corresponding jacobian into account. The result of all this is that
the effective action $\Gamma_{eff}[T]$ is indeed of the form
$Z_k c \Gamma^{(0)}_{eff}[Z_T T]$, with $Z_k$ and $Z_T$ given by
(\ref{grav}).

In the next section we perform this calculation for $W_3$
gravity. Ordinary gravity can be treated in the same way, and we
leave the detailed calculation for ordinary gravity to the
reader. The induced and effective action for $W_3$ action are
defined similarly as for ordinary gravity. The induced action
depends, besides on $\mu$, on an extra field $\nu$ (sometimes
called $B_{\bar{z}\bar{z}\bar{z}}$), that couples to the $W_3$
field. A difference with ordinary gravity is that
the explicit form of $\Gamma_c[\mu,\nu]$ is not known; for $W_3$
gravity $\Gamma_c[\mu,\nu]$ has an expansion in $1/c$, and
is known up to three loops \cite{sb-w3ind}. For $W_3$ gravity we
start with a constrained $SL(3,\re)$ WZW model. It is known that
imposing certain constraints on an $SL(3,\re)$ current algebra
reduces the current algebra to a $W_3$ algebra
\cite{beroog,drisok}. The fields $T(g)$ and $W(g)$ that
couple to the $W_3$ gauge fields $\mu,\nu$ are the generators of
this $W_3$ algebra. Furthermore,
$T(g)$ and $W(g)$ can be chosen such as to preserve the
gauge invariance of the constrained WZW model. This enables us
to perform a BRST quantization of the model. Because the BRST
operator is nilpotent only on-shell, we need the
Batalin-Vilkovisky quantization procedure to compute the quantum
action.

To complete the computation, we need the jacobian for the change
of variables from $g$ to $g^{-1}\dif g$. This is a rather
difficult point, of which our understanding is incomplete. This
point is discussed in section 4, where it is shown that knowing
this jacobian is equivalent to knowing the effective action for
ordinary WZW theory. Using the ansatz that this effective action
is proportional, up to multiplicative renormalizations, to the
WZW action, we then complete the calculation of the effective
action of $W_3$ gravity. The result agrees with
one-loop calculations\cite{grini,vannie} for $W_3$ gravity if
the multiplicative renormalizations of the WZW model agree
with the one-loop calculations for the WZW model
\cite{pol4,wreview3}.
The resulting effective action for $W_3$ gravity
is proportional to a
constrained WZW model, as conjectured in \cite{vannie}.
Thus, $W_3$ gravity can be seen as an
example of completely integrable nonlocal field theory. The
crucial ingredient in establishing this integrability is
demanding BRST invariance at the quantum level.
The result also shows that the level of
the $SL(3,\re)$ current algebra in $W_3$ gravity is given by a
KPZ-like formula as proposed in \cite{matsuo,beroog}.
We would like to stress that none of these conclusions holds if
the effective action of the WZW model is not simply proportional
to a WZW action.

Actually, Knizhnik, Polyakov and Zamolodchikov derived their
result for the level of the $SL(2,\re)$ current algebra in
gravity by an analysis of the gauge fixing of the covariant
induced action for gravity \cite{kpz}. This procedure is closely
related to the one used here, and can be generalized to $W_3$
gravity by gauge fixing the covariant action \cite{jj,jj2}
for $W_3$ gravity.
The advantage of this approach is that it makes
the $SL(3,\re)$ current algebra structure in $W_3$ gravity very
clear. The disadvantage is that it is difficult to extract all
order results from it, because the covariant action is only
known to lowest order in $1/c$. This 'KPZ' approach to $W_3$
gravity will be discussed in a separate paper \cite{jj3}.

\newsection{The Induced Action of $W_3$ Gravity}

We start with the action for a constrained $SL(3,\re)$ WZW
model. The action
is given by \cite{beroog}
\be \label{eq:WZW1}
S_1=kS^-_{WZW}(g)+\aksie\ (\bar{A}^1 (J_1-\xi)+\bar{A}^2 (J_2-\xi)
+ \bar{A}^3 J_3 ),
\ee
where the current ${\cal J}=k g^{-1}\dif g$ is parametrized by
\be \label{def:j}
{\cal J}=\mat{H_0}{J_1}{J_3}{K_1}{H_1-H_0}{J_2}{K_3}{K_2}{-H_1}.
\ee
The action consists of a WZW model at level $k$, and
three gauge fields $\bar{A}^i$ ($i=1\ldots3$), that play the role
of lagrange multipliers; $\xi$ is an arbitrary parameter different
from zero, that is usually taken to be equal to one. It is well known
that imposing the constraints $J_1=J_2=\xi,J_3=0$ on an $SL(3,\re)$
current algbra reduces the current algebra to a $W_3$ algebra
\cite{beroog,drisok}.
The WZW actions $S^{\pm}_{WZW}$ are given by
\be \label{deff}
S^{\pm}_{WZW}(g)=\deel{1}{2\pi}\int_{\Sigma}d^2z\,\tr(g^{-1}\dif
gg^{-1}\dbar g)\pm \deel{1}{6\pi}\int_B \tr(g^{-1}dg)^3,
\ee
and satisfy the following Polyakov--Wiegmann identities \cite{powie}:
\ba \label{idd}
S_{WZW}^{+}(gh) &=& S_{WZW}^{+}(g)+S_{WZW}^{+}(h)
+\accc{g^{-1}\dbar g\dif h h^{-1}}, \nonu
S_{WZW}^{-}(gh) &=& S_{WZW}^{-}(g)+S_{WZW}^{-}(h)
+\accc{g^{-1}\dif g \dbar h h^{-1}}.
\ea
The action (\ref{eq:WZW1}) has an invariance under the gauge
transformations generated by the subgroup $N^-$ of lower
triangular matrices. Explicitly, the action (\ref{eq:WZW1}) is
invariant under $\delta_{\eps}{\cal J}=k\dif\eps+[{\cal J},\eps]$ (or
$\delta_{\eps}g=g\eps$) and
$\delta_{\eps}\bar{A}^1=-\dbar\eps_1$,
$\delta_{\eps}\bar{A}^2=-\dbar\eps_2$, and
$\delta_{\eps}\bar{A}^3=-\dbar\eps_3+\bar{A}^2\eps_1-\bar{A}^1
\eps_2$, where
\be \label{def:eps}
\eps=\mat{0}{0}{0}{\eps_1}{0}{0}{\eps_3}{\eps_2}{0}.
\ee
As explained in the introduction we intend to couple this theory
to the $W_3$ gauge fields $\mu,\nu$ by adding a term
$\int \mu T({\cal J})+\int\nu
W({\cal J})$ to the action, while preserving the
gauge invariance.
To find $T({\cal J})$ and $W({\cal J})$ one uses
the fact that there is a unique gauge transformation given by a
lower triangular matrix $n$ with ones on the diagonal, such that
\be \label{eq:miura}
\mat{0}{\xi}{0}{T({\cal J})/2\xi}{0}{\xi}{W({\cal J})/\xi^2}{T({\cal
J})/2\xi}{0}=n^{-1}
\mat{H_0}{\xi}{0}{K_1}{H_1-H_0}{\xi}{K_3}{K_2}{-H_1}n +
k n^{-1}\dif n.
\ee
The factors $1/2\xi$ and $1/\xi^2$ have been included
for later convenience.
The polynomials $T({\cal J})$ and $W({\cal J})$ are invariant under
$N^-$ gauge transformations of the constrained current ${\cal
J}_{constr}={\cal J}|_{
J_1=\xi,J_2=\xi,J_3=0}$
that appears in (\ref{eq:miura}): if we perform the $N^-$ gauge
transformation ${\cal J}'_{constr}=m^{-1} {\cal J}_{constr} m + k
m^{-1}\dif m$, then the unique lower triangular matrix $n$ in
(\ref{eq:miura}) that brings ${\cal J}_{constr}'$ in the right form is
given by $n'=m^{-1}n$. Because $n^{-1}{\cal J}_{constr} n + k n^{-1}
\dif n=n'^{-1} {\cal J}'_{constr} n' + k n'^{-1} \dif n'$, the left hand
side of (\ref{eq:miura}) does not change under this gauge
transformation, and $T({\cal J})$
and $W({\cal J})$ are gauge invariant. Under
a gauge transformation of the full current ${\cal J}$, $T({\cal
J})$ and
$W({\cal J})$ are only invariant up to terms proportional to
${\cal J}-{\cal J}_{constr}$.
Therefore, if we add $\int \mu
T({\cal J}) +\int \nu W({\cal J})$
to the action (\ref{eq:WZW1}), the action
is $N^-$ invariant up to terms proportional to the
constraints. It is possible,
by modifying the transformation rules for
$\bar{A}^i$, to make the action exactly $N^-$ invariant.
The expressions $T({\cal J})$ and $W({\cal J})$
are the so-called gauge
invariant polynomials on the constrained phase space
\cite{dublinann}. In terms of
classical hamiltonian reduction, it are precisely these
polynomials that survive the hamiltonian reduction of the WZW
theory and it is known \cite{dublinann} that they
generate an algebra that is isomorphic to the
classical $W_3$ algebra.
Of course, if $W$ and $T$ are gauge invariant polynomials on the
constrained phase space, so are $W+\alpha\dif T$ and $T$. The $T$
and $W$ we take, as defined by (\ref{eq:miura}), correspond to
a particular basis choice known as the 'highest
weight gauge' \cite{dublinann},
which guarantees that $W$ will transform as a spin three field.

If we compute $T({\cal J})$ and $W({\cal J})$ from
(\ref{eq:miura}) and add these to the action (\ref{eq:WZW1}),
the resulting action $S_2(\bar{A},g,\mu,\nu)$ reads
\ba
S_2
& = & kS^-_{WZW}(g)+\aksie\ (\bar{A}^1 (J_1-\xi)+\bar{A}^2 (J_2-\xi)
+ \bar{A}^3 J_3 ) \nonu
& & + \aksies{N_T} \mu(H_0^2-H_0 H_1 +H_1^2 + \xi(K_1+K_2) -k\dif(
H_0+H_1)) \nonu
& & + \aksies{N_W} \nu (H_0^2 H_1 - H_0 H_1^2 +\xi (H_1 K_1-H_0
K_2) + \xi^2 K_3 + \hf k \xi \dif (K_2-K_1) \nonu
& & + \hf k^2 \dif^2 ( H_0 - H_1) + k (-H_0\dif H_0 + H_1 \dif
H_1 +\hf H_0 \dif H_1 - \hf H_1 \dif H_0)),
\label{eq:WZW2}
\ea
where we have introduced two normalization factors $N_T$ and $N_W$.
As explained above, the $N^-$ transformations that leave this action
invariant are still given by $\delta_{\eps}
{\cal J}=k\dif\eps +[{\cal J},\eps]$ for the current,
while for $\bar{A}^i$ they are
extended to
\ba
\delta_{\eps}\bar{A}^3 & = & -\dbar\eps_3+\bar{A}^2\eps_1-\bar{A}^1
\eps_2
+\eps_3 (-N_T (\mu (H_0+H_1)+2 k \dif \mu) \nonu
& & +N_W(
\nu(-H_0^2+H_1^2+\xi( K_2 -K_1) +k\dif(H_0-H_1))
+\deel{k}{2}\dif\nu (H_1-H_0))), \nonu
\delta_{\eps}\bar{A}^2 & = & -\dbar\eps_2+N_T \mu(\eps_2 (H_0-2
H_1) -\xi\eps_3)-k N_T \eps_2 \dif\mu  \nonu & & +
N_W \nu (-\eps_2 H_0
(H_0-2 H_1)-\xi\eps_3 H_1 +\xi\eps_2 K_1 + k\eps_2 \dif H_0)
\nonu
& & +\deel{k}{2} N_W (\eps_2 \dif \nu (H_0 + 2 H_1) - \xi \eps_3 \dif
\nu) +
 \deel{k^2}{2} N_W \eps_2 \dif^2 \nu, \nonu
\delta_{\eps}\bar{A}^1 & = & -\dbar\eps_1+N_T \mu(\eps_1 (2 H_0-
H_1) +\xi\eps_3)-k N_T \eps_1 \dif\mu  \nonu
& & - N_W \nu (-\eps_1 H_1
(H_1-2 H_0)+\xi\eps_3 H_0 -\xi\eps_1 K_2 + k\eps_1 \dif H_1) \nonu
& & -
\deel{k}{2} N_W (\eps_1 \dif \nu (2 H_0 +  H_1) + \xi \eps_3 \dif
\nu)
 - \deel{k^2}{2} N_W \eps_1 \dif^2 \nu. \label{eq:trafo1}
\ea
As a generalization of (\ref{def:sindgr2})
we consider the functional $S_{ind}(\mu,\nu)$ defined by
\be \label{eq:defind}
e^{-S_{ind}(\mu,\nu)}=\int\frac{{\cal D}\bar{A} {\cal D}g}{\mbox{\rm
gauge volume}}
e^{-S_2(\bar{A},g,\mu,\nu)}.
\ee
We shall prove shortly, that, with an appropriate choice
of $N_W$ and $N_T$, the
induced action $S_{ind}(\mu,\nu)$ is equal to the induced action
for $W_3$ gravity, to all orders in $1/c$. The induced action
$\gam_c[\mu,\nu]$ for $W_3$ gravity is defined by
(cf. (\ref{def:sindgr}))
\be \label{def:sind}
e^{-\Gamma_c[\mu,\nu]}=\left\langle e^{-\aksie\ (\mu T + \nu W)}
\right\rangle_{OPE},
\ee
where the right hand side is computed using the operator product
expansions of the $W_3$ algebra. Thus, in order to prove that
$S_{ind}(\mu,\nu)=\gam_c[\mu,\nu]$, we need to verify that upon
quantizing $g$ and $\bar{A}$ the fields
\be \label{def:indf}
T_{ind}=\pi\vars{S_2}{\mu}, \,\,\,\,\,\,\,\,\,
W_{ind}=\pi\vars{S_2}{\nu},
\ee
generate a quantum $W_3$ algebra. This fixes the values of
$N_W$ and $N_T$.

The quantization of $S_2(g,\bar{A},\mu,\nu)$ is most easily
performed using BRST quantization (cf. \cite{beroog}).
The BRST transformation
rules for $g$ and $\bar{A}$ are defined by replacing the
parameters $\eps_i$ of the gauge transformations
$\delta_{\eps}g=g\eps$ and (\ref{eq:trafo1}) by anti-commuting
ghosts $c_i$. We denote these transformation rules by $\delta_B
g$ and $\delta_B \bar{A}$. The BRST transformation rules for the
ghosts read $\delta_B c_1=\delta_B c_2=0$ and $\delta_B c_3=c_1
c_2$. However, due to the extra terms we added to the $\bar{A}$
transformation rules in (\ref{eq:trafo1}), the BRST operator
$\delta_B$ no longer satisfies $\delta_B^2=0$. It only satisfies
$\delta_B^2=0$ when we use the $\bar{A}$ equations of motion.
In such a case a proper quantization and BRST gauge fixing of
the theory requires that we use the Batalin-Vilkovisky
formalism \cite{batvi}.

For all fields in the theory we introduce antifields
($\bar{A}_i^{\ast}$, $g^{\ast}$ and
$c^{i\ast}$) with opposite statistics.
Because $\delta_B^2=0$ only on-shell, we
typically need to include terms that are quadratic in the ghosts
$c_{\alpha}$
and in the anti-fields to find a solution to the master equation.
Because only $\delta_B^2 \bar{A}_i \neq 0$, the only terms
quadratic in the antighosts that are needed are terms quadratic
in $\bar{A}_i^{\ast}$. Furthermore, if we compute $\delta_B^2
\bar{A}_i$, we find that each term in the answer contains at
most one derivative, and that the answer is proportional to the
$\bar{A}_i$ equations of motion.
This leads us to write down the following ansatz for the minimal
solution to the master equation
\ba
S_{min} & = & S_2+\int\bar{A}_i^{\ast}\delta_B\bar{A}^i+\int
g^{\ast}\delta_B g
-\int c^{3\ast}c_1c_2
+\int \bar{A}^{\ast}_i \bar{A}^{\ast}_j E^{ij,\alpha\beta}c_{\alpha}
c_{\beta} \nonu & &
+\int
\bar{A}^{\ast}_i \bar{A}^{\ast}_j F^{ij,\alpha\beta}c_{\alpha} \dif
c_{\beta}
+\int
\bar{A}^{\ast}_i \dif \bar{A}^{\ast}_j G^{ij,\alpha\beta}c_{\alpha}
c_{\beta}.
\label{eq:smin}
\ea
If we denote by $\phi^I$ the set of fields $(\bar{A}^i,g,c_i)$
and by $\phi^{\ast}_I$ the corresponding set of anti-fields
$(\bar{A}^{\ast}_i,g^{\ast},c^{i\ast})$, then the master
equation reads $(S_{min},S_{min})=0$, where
\be \label{def:bra}
(P,Q)= \frac{\stackrel{\leftarrow}{\dif P}}{\dif\phi^I}
\frac{\stackrel{\rightarrow}{\dif Q}}{\dif\phi^{\ast}_I}-
\frac{\stackrel{\leftarrow}{\dif P}}{\dif\phi^{\ast}_I}
\frac{\stackrel{\rightarrow}{\dif Q}}{\dif\phi^I}.
\ee
Working out the master equation for (\ref{eq:smin}) yields,
among others, the equation
\ba
\delta_B^2(\bar{A}^k) & = & \vars{S_2}{\bar{A}^j} \left((2E^{jk,\alpha
\beta}-\dif G^{jk,\alpha\beta})
c_{\alpha} c_{\beta} + (2F^{jk,\alpha\beta}
-G^{jk,\alpha\beta} + G^{jk,\beta\alpha})c_{\alpha}
\dif c_{\beta} \right)\nonu
& & -\dif\left(\vars{S_2}{\bar{A}^j}\right)
(G^{jk,\alpha\beta}+G^{kj,\alpha\beta})c_{\alpha}c_{\beta}.
\label{eq:nil}
\ea
{}From this one can compute the tensors $E$, $F$ and $G$. The
components of these tensors either vanish, or can be determined
from the following relations
\ba
E^{jk,\alpha\beta} & = &
=-E^{kj,\alpha\beta}=-E^{jk,\beta\alpha}, \nonu
E^{12,12} & = & \deel{1}{4} (N_T \mu - 2 N_W H_0 \nu + 2 N_W H_1
\nu), \nonu
E^{13,13} & = & \deel{1}{4} (-N_T \mu -2 N_W H_0 \nu - k N_W
\dif\nu), \nonu
E^{23,23} & = & \deel{1}{4} (-N_T \mu + 2 N_W H_1 \nu + k N_W
\dif\nu), \nonu
G^{jk,\alpha\beta} & = & G^{kj,\alpha\beta}=-G^{jk,\beta\alpha},
\nonu
G^{12,12} & = & G^{13,13}=-G^{23,23}=-\deel{k}{4} N_W \nu, \nonu
F^{jk,\alpha\beta} & = & -F^{kj,\alpha\beta}=F^{jk,\beta\alpha},
\nonu
F^{12,12} & = & F^{13,13}=-F^{23,23}=-\deel{k}{4} N_W \nu.
\label{eq:efg}
\ea
If we substitute this back into (\ref{eq:smin}), we find that
the master equation is satisfied. The full quantum action is
given by
\be \label{eq:squa}
S_q=S_{min}-\aksie (b_1^{\ast}B_1 +b_2^{\ast}B_2+b_3^{\ast}B_3),
\ee
where $b_i^{\ast}$ are the anti-fields for the anti-ghosts
$b_i$, and the $B_i$ are Lagrange multipliers, also known as the
Nakanishi-Lautrup fields, that will impose the gauge
condition. The gauge fixing is done by replacing
the antifields $\phi^{\ast}$ by
$\dif\Psi/\dif\phi$ in the full quantum action (\ref{eq:squa}),
where $\Psi$, the gauge fermion,
represents a particular gauge
choice. We will choose
\be \label{def:psi}
\Psi=\int d^2 z \, (b_1\bar{A}^1 + b_2 \bar{A}^2 + b_3 \bar{A}^3),
\ee
so that we put $c^{i\ast}=g^{\ast}=0$, $\bar{A}^{\ast}_i=b_i$
and $b^{\ast}_i=\bar{A}^i$ in (\ref{eq:squa}). The resulting
gauge fixed action is off-shell BRST invariant under the BRST
transformations
\be \label{def:brst}
\delta_B' \phi^I=- \left. \frac{\stackrel{\leftarrow}{\dif S_q}}{\dif
\phi^{\ast}_I} \right|_{\phi_I^{\ast}=\dif\Psi/\dif\phi^I}.
\ee
Note that the transformation rules for $\bar{A}$ with respect to
$\delta_B'$ are different from those with respect to $\delta_B$,
but we are going to integrate out the $\bar{A}$, we do not give
those (lengthy) transformation rules here. The gauge fixed
action we have obtained can be written in a form that is remarkably
similar to  (\ref{eq:WZW2}),
\ba
S_{gf} & = &
kS^-_{WZW}(g)-\aksie\ (b_1\dbar c_1 + b_2 \dbar c_2 + b_3 \dbar
c_3) \nonu
& & +\aksie\ (\bar{A}^1 (\hat{J}_1-\xi-B_1)+
\bar{A}^2 (\hat{J}_2-\xi-B_2)
+ \bar{A}^3 (\hat{J}_3-B_3) ) \nonu
& & + \aksies{N_T} \mu(\hat{H}_0^2-\hat{H}_0 \hat{H}_1
+\hat{H}_1^2 + \xi(\hat{K}_1+\hat{K}_2) -k\dif(
\hat{H}_0+\hat{H}_1)) \nonu
& & + \aksies{N_W} \nu (\hat{H}_0^2 \hat{H}_1
- \hat{H}_0 \hat{H}_1^2 +\xi (\hat{H}_1 \hat{K}_1-\hat{H}_0
\hat{K}_2) + \xi^2 \hat{K}_3
+ \hf k \xi \dif (\hat{K}_2-\hat{K}_1) \nonu
& & + \hf k^2 \dif^2 ( \hat{H}_0 - \hat{H}_1)
+ k (-\hat{H}_0\dif \hat{H}_0 + \hat{H}_1 \dif
\hat{H}_1 +\hf \hat{H}_0 \dif \hat{H}_1
- \hf \hat{H}_1 \dif \hat{H}_0)),
\label{eq:WZW3}
\ea
where the hatted currents are the components of an $SL(3,\re)$
valued object
$\hat{\cal J}$
and are defined by
\ba
& \hat{J}_1=J_1+c_2 b_3, \qquad\;\;\; \hat{J}_2=J_2-c_1
b_3,  \qquad\;\;\; \hat{J}_3=J_3, & \nonu
& \hat{H}_0=H_0+c_1 b_1 +c_3
b_3,  \qquad\,\,\, \hat{H}_1=H_1+c_2b_2+c_3b_3, & \nonu
& \hat{K}_1=K_1+c_3 b_2,  \qquad \hat{K}_2=K_2-c_3
b_1,  \qquad \hat{K}_3=K_3. &
\label{def:hat}
\ea
A simple way to define these hatted quantities is by means of
the following expression
\be \label{def:hat2}
 \hat{\cal J}={\cal J} -
 \left[
\mat{0}{0}{0}{c_1}{0}{0}{c_3}{c_2}{0}  \!\! ,  \!\!
\mat{0}{b_1}{b_3}{0}{0}{b_2}{0}{0}{0} \right]_+,
\ee
where $[,]_+$ denotes an anticommutator. The meaning of these
hatted currents becomes clear once we integrate out
$B_i$ from the gauge fixed action $S_{gf}$, giving
\ba
S_{gf2} & = &
kS^-_{WZW}(g)-\aksie\ (b_1\dbar c_1 + b_2 \dbar c_2 + b_3 \dbar
c_3) \nonu
& & + \aksies{N_T} \mu T(\hat{{\cal J}})
 + \aksies{N_W} \nu W(\hat{{\cal J}}).
\label{eq:WZW4}
\ea
The BRST transformation rules for the anti-ghosts $b_i$ now read
\ba
\delta_B b_1 & = & J_1-\xi + c_2 b_3, \nonu
\delta_B b_2 & = & J_2-\xi - c_1 b_3, \nonu
\delta_B b_3 & = & J_3.
\label{eq:delb}
\ea
If we compare the BRST transformation rules of $H_i$ and $K_i$
with those for $\hat{H}_i$ and $\hat{K}_i$, we see that
the transformation rules for
$\hat{H}_i$ and $\hat{K}_i$ can be obtained
from those for $H_i$ and $K_i$ by
replacing $J_1$ and $J_2$ by $\xi$ and $J_3$ by $0$,
and $H_i$ and $K_i$ by their hatted counterparts. The BRST
transformation rules for $\hat{H}_i$ and $\hat{K}_i$ are
therefore determined by the way the constrained current behaves
under $N^-$ gauge transformations, whereas the transformation
rules for $H_i$ and $K_i$ were determined by the way in which
the unconstrained current transformed under gauge transformations.
Because $T({\cal J})$ and $W({\cal J})$
were constructed in such a way as to
be exactly invariant under $N^-$ gauge transformations of the
constrained current, this automatically implies that
$T(\hat{{\cal J}})$ and $W(\hat{{\cal J}})$
must be BRST invariant. The same
procedure also works for constrained $SL(N,\re)$ models for
arbitrary $N$. Instead of going through all the details of the
Batalin-Vilkovisky procedure, one simply constructs the gauge
invariant polynomials on the constrained phase space and then
replaces currents by hatted currents, using the obvious
generalizations of (\ref{eq:miura}) and (\ref{def:hat2}),
to construct the BRST invariant gauge fixed action. Actually,
this procedure is nothing but a classical version of the quantum
hamiltonian reduction studied in \cite{feifre}. In this
approach, one computes the cohomology of the BRST operator
generating the BRST transformations of the gauge fixed action
(\ref{eq:WZW4}) with $\mu=\nu=0$
on the space of polynomials in the currents and
the ghosts. The result is that the BRST cohomology is generated
by quantum versions $T_q$ and $W_q$ of
$T(\hat{{\cal J}})$ and $W(\hat{{\cal J}})$,
which are given by (\ref{def:twq})
\footnote{In \cite{feifre}, it is shown that the BRST cohomology
is isomorphic to the algebra generated by $T_q$ and $W_q$ with
$\hat{K}_i=0$. The actual BRST representatives of the cohomology
were not constructed in \cite{feifre}, but one can show that they
are given by (\ref{def:twq}). Indeed, the algebra generated by
$T_q$ and $W_q$ does not change if one puts $\hat{K}_i=0$ in
(\ref{def:twq})}.
Here we see that this quantum BRST cohomology is a
quantization of the space of gauge invariant polynomials on the
constrained phase space of the classical theory.

The gauge fixed action (\ref{eq:WZW4}) consists of a WZW model
and of three $b,c$ systems. Because we know how to quantize
these \cite{fms,kz}, we can now try to find out whether
$S_{ind}(\mu,\nu)$ defined in (\ref{eq:defind}) really is the
induced action for $W_3$ gravity to all orders. As we explained
previously, for this we need that $N_T T(\hat{{\cal J}})$ and $N_W
W(\hat{{\cal J}})$ generate, at the quantum level, a $W_3$ algebra. The
easiest way to check this is to use operator product expansions,
because we know what these are for the current ${\cal J}$ and for the
ghosts. However, it is not clear how we should replace
$T(\hat{{\cal J}})$ and $W(\hat{{\cal J}})$
by normal ordered expressions
involving the currents and the ghosts. For instance, using the
normal ordering prescription of \cite{sander}, it is not true
that $(H_0 K_2)=(K_2 H_0)$. Two normal
orderings of a product of a certain number of currents always
differ by terms that contain fewer currents than the original
product.
This indicates that the coefficient in front
of the term with the largest number of currents is the same,
both for the classical expressions $T(\hat{{\cal J}}),W(\hat{{\cal
J}})$
and their normal ordered versions $T_q({\cal J},\mbox{\rm ghosts}),
W_q({\cal J},\mbox{\rm ghosts})$. To obtain the full expressions for
$T_q$ and $W_q$, we need some extra principle that tells us how
to do this. The extra principle we choose is that of BRST
invariance. As $T(\hat{{\cal J}}),W(\hat{{\cal J}})$
were classically BRST
invariant, we require that $T_q,W_q$ are quantum BRST
invariant. Together with the requirement that the coefficients
for the terms with the largest number of currents do not change,
this will completely fix the form of $T_q$ and $W_q$.
Classically, the BRST charge is given by
\be \label{def:q}
Q=\oint \frac{dz}{2\pi i} (c_1(\xi-J_1)+c_2(\xi-J_2)-c_3 J_3-c_1
c_2 b_3 ).
\ee
The quantum BRST operator is given by the same
expression, with products of fields
replaced by normal ordered products. Notice
that there is no normal ordering ambiguity in the
definition of $Q$.
The OPE's of the ghosts and the currents are given
by
\ba
c_i(z) b_j(w) & = & \frac{-\delta_{ij}}{(z-w)},
\nonu
{\cal J}_a(z) {\cal J}_b(w) & = & \frac{-k \eta_{ab}}{(z-w)^2}
+\frac{-f_{ab}^{\,\,\,\,\,c}{\cal J}_c(w)}{(z-w)},
\label{def:ope}
\ea
where we decomposed the current ${\cal J}={\cal J}_a T^a$,
$\eta^{ab}=\tr(T^a T^b)$, $\eta_{ab}$ is the
inverse of $\eta^{ab}$, $f^{ab}_{\,\,\,\,\,
c}\,T^c=[T^a,T^b]$, and indices are raised and
lowered using $\eta^{ab}$.
It is now a straightforward computation to show
that the fields
\ba
T_q & = & N_T ((\hat{H}_0\hat{H}_0)-(\hat{H}_0 \hat{H}_1)
+(\hat{H}_1\hat{H}_1) + \xi(\hat{K}_1+\hat{K}_2) -(k-2)\dif(
\hat{H}_0+\hat{H}_1)), \nonu
W_q & = & N_W ( (\hat{H}_0(\hat{H}_0\hat{H}_1))
- (\hat{H}_0 (\hat{H}_1\hat{H}_1))
+\xi ((\hat{H}_1 \hat{K}_1)-(\hat{H}_0
\hat{K}_2)) + \xi^2 \hat{K}_3 \nonu
& & + \hf (k-2) \xi \dif (\hat{K}_2-\hat{K}_1)
+ \hf (k-2)^2 \dif^2 ( \hat{H}_0 - \hat{H}_1) \nonu
& & + (k-2) (-(\hat{H}_0\dif \hat{H}_0) + (\hat{H}_1 \dif
\hat{H}_1) +\hf (\hat{H}_0 \dif \hat{H}_1 )
- \hf (\hat{H}_1 \dif \hat{H}_0))),
\label{def:twq}
\ea
form a quantum $W_3$ algebra, with central charge
\be \label{def:c}
c=50+24\left((k-3) + \frac{1}{(k-3)} \right).
\ee
Here, the hatted fields are still given by
(\ref{def:hat}), and have the OPE's
\be \label{def:hatope}
\hat{{\cal J}}_a(z) \hat{{\cal J}}_b(w)
=  \frac{-(k-3) \eta_{ab}}{(z-w)^2}
+\frac{-f_{ab}^{\,\,\,\,\,c}\hat{{\cal J}}_c(w)}{(z-w)},
\ee
for $\hat{{\cal J}}_a,\hat{{\cal J}}_b \in \{
\hat{H}_i,\hat{K}_i \}$.
The normalization constants $N_T$ and $N_W$ must be
equal to
\ba
N_T & = & \frac{-1}{k-3}, \nonu
N_W & = & \left(\frac{-6}{15k^4-146k^3+519k^2-792 k
+432}\right)^{\frac{1}{2}} = \left(
\frac{-48}{(k-3)^3 (5c+22)} \right)^{\frac{1}{2}}.
\label{def:norm}
\ea
This shows that $S_{ind}(\mu,\nu)$ is indeed the all order
induced action for $W_3$ gravity, where $c$ is related to $k$
via (\ref{def:c}) and $N_T$ and $N_W$ must be chosen according to
(\ref{def:norm}). The constant $\xi$ can be chosen arbitrarily.

To summarize, we have shown that the constrained WZW model can
be coupled to the $W_3$ gauge fields in such a way that the
resulting induced action for the $W_3$ gauge fields is precisely
the all order (chiral) induced action for $W_3$ gravity.

\newsection{The Effective Action of $W_3$ Gravity}

The effective action for $W_3$ gravity is obtained by quantizing the
induced action, and is defined by the following path integral
(cf. (\ref{def:seffgr}))
\be \label{def:seff}
e^{-\Gamma_{eff}[T,W]}=\int \frac{{\cal D} g {\cal D} \bar{A}}{\mbox{
\rm gauge volume}}{\cal D} \mu
{\cal D} \nu e^{\aksie(\mu T+\nu W)-S_2(g,\bar{A},\mu,\nu)}  ,
\ee
where $S_2$ is the action (\ref{eq:WZW2}). In the previous
section we performed a BRST quantization of
$S_2(g,\bar{A},\mu,\nu)$, by gauge fixing $\bar{A}^i=0$. This is
a convenient gauge condition for proving that the induced action
for $\mu$ and $\nu$ is the same as the induced action for $W_3$
gravity, but not for the
computation of the effective action. Therefore, we will use a
different gauge here, namely $H_0=H_1=K_1-K_2=0$. Because the
BRST operator $\delta_B$ satisfies $\delta^2_B H_0=\delta^2_B
H_1=\delta^2_B(K_1-K_2)=0$, there is no need to use
Batalin-Vilkovisky quantization here.
Under gauge transformations
$H_0,H_1$ and $K_1-K_2$ transform as
\ba
\delta_{\eps} H_0 & = & J_1 \eps_1 + J_3 \eps_3, \nonu
\delta_{\eps} H_1 & = & J_2 \eps_2 + J_3 \eps_3, \nonu
\delta_{\eps} (K_1-K_2) & = & (H_1-2H_0)\eps_1 + (H_0-2H_1)\eps_2 +
(J_2-J_1)\eps_3 + k \dif (\eps_1-\eps_2).
\label{eq:trafo2}
\ea
This shows that gauge
fixing $H_0=H_1=K_1-K_2=0$ produces a Faddeev-Popov
contribution to the path integral which is equal to
\be \label{eq:fp}
\int \! {\cal D}\beta_1 {\cal D} \gamma_1
 {\cal D}\beta_2 {\cal D} \gamma_2
 {\cal D}\beta_3 {\cal D} \gamma_3 \exp
 \left(-\aksie (\xi \beta_1
 \gamma_1 + \xi \beta_2 \gamma_2 + 2 \xi \beta_3
 \gamma_3 + k \beta_3 \dif (\gamma_1-\gamma_2)) \right),
 \ee
 where we put $J_1=J_2=\xi$ and $J_3=0$, which can be done safely after
 performing the $\bar{A}$ integration.
 It is clear that (\ref{eq:fp}) is just
 some numerical factor, and we will ignore this factor.
 Then we can remove the volume of the gauge group
in (\ref{def:seff}) by inserting the combination
 $\delta(H_0)\delta(H_1)\delta(K_1-K_2)$ into the path integral.
The $\bar{A}$ and $\mu,\nu$
integrations yield
five more delta function insertions in the path integral.
Altogether this shows that
 \ba
 e^{-\Gamma_{eff}[T,W]} & = & \int {\cal D} g
 \delta(J_1-\xi) \delta(J_2-\xi)
 \delta(J_3) \delta(H_0) \delta(H_1) \delta (K_1-K_2) \nonu & &
 \delta(T-N_T\xi(K_1+K_2))
 \delta(W-N_W\xi^2K_3) e^{-kS_{WZW} (g)}.
 \label{eq:seff2}
 \ea
 It seems that we are already done,
 as the delta functions absorb all the
 degrees of freedom, and that we are left with a
 constrained WZW model. However,
 before we can integrate out the delta functions,
 we must first change variables from $g$
 to $g^{-1}\dif g$, and compute the corresponding jacobian.
 This change of variables is a
 rather tricky point, which we now discuss in some detail.

\newsection{The Effective Action of the WZW Model}

It is generally believed \cite{powie,pol4},
that the jacobian corresponding
to the change of variables from $A_z=g^{-1}\dif g$ to $g$
leads to\footnote{The symbols $A_z$ and $J_{\bar{z}}$
used in this section should not be
confused with $\bar{A}^i$ and $J_i$ used in the previous
sections.}
\be \label{eq:jac}
 {\cal D} A_z = exp(-2 h_G
S^-_{WZW}(g) )\, {\cal D}g ,
\ee
where $h_G$ is the dual
Coxeter number of the group under consideration.
The computation of this jacobian proceeds by noticing that $\delta
A_z=\dif_{A_z}(g^{-1}\delta g)$, so that the jacobian is equal to
$\det(\dif_{A_z})$, and then by writing this determinant as the path
integral $\int{\cal D}\psi {\cal D} \bar{\psi}\exp(-\int \bar{\psi}
\dif_{A_z} \psi)$, where $\psi,\bar{\psi}$ are fermions transforming in
the adjoint representation of the group. Finally, one can derive a Ward
identity for this fermionic path integral and show that the solution to
this Ward identity is indeed given by (\ref{eq:jac}).

Actually, (\ref{eq:jac}) is in disagreement with one-loop
calculations for
the WZW model \cite{pol4,wreview3}.
If (\ref{eq:jac}) were true, then one could easily compute
the effective action for the WZW model to all orders: first, we compute
the generating functional of connected diagrams $G[J_{\bar{z}}]$,
given by
\be \label{def:sindWZW}
\exp{-G[J_{\bar{z}}]}=\int {\cal D}
A_z \exp(-kS^-_{WZW}(A_z) + \aksie \tr (A_z
J_{\bar{z}})).
\ee
If we change variables from $A_z$ to $g$ with $A_z=g^{-1}\dif g$, and
parametrize $J_{\bar{z}}$ by $J_{\bar{z}}=-(k+2h_G)\dbar h h^{-1}$,
we can use the
Polyakov-Wiegmann identity (\ref{idd})
to write the right-hand side of (\ref{def:sindWZW}) as
\be \label{eq:sindWZW2}
\int {\cal D} g \exp( -(k+2h_G)S^-_{WZW}(gh)+(k+2h_G)S^-_{WZW}(h)).
\ee
We can safely replace the
variable $g$ by $g'=gh^{-1}$, because this does
not change the measure ${\cal D}g$, and we see that if we ignore an
infinite factor, the generating functional
$G[J_{\bar{z}}]=-(k+2h_G)S^-_{WZW}(h)$.
The effective action $S_{eff}(A_z)$ is the
Legendre transform of $G[J_{\bar{z}}]$,
\ba \label{def:seffWZW}
S_{eff}(A_z) & = & \min_{J_{\bar{z}}}
\left(-G[J_{\bar{z}}]-\aksie \tr (A_z J_{\bar{z}}) \right)
\nonu & =  & \min_{h}
\left((k+2h_G)S^-_{WZW}(h) + \aksies{(k+2h_G)} \tr (A_z \dbar h h^{-1})
\right)
\nonu & =  & \min_{h}
\left((k+2h_G)S^-_{WZW}(h^{-1}) -
\aksies{(k+2h_G)} \tr (A_z   h^{-1}\dbar h)
\right).
\ea
The extremum is attained for $A_z=h^{-1}\dif h$, and we find that the
effective action is simply
\be \label{eq:seff3}
S_{eff}(A_z)=-(k+2h_G)S^-_{WZW}(A_z).
\ee

On the other hand, one can also perform a one-loop computation
of the effective action \cite{pol4,wreview3},
and check the above result.
In (\ref{def:sindWZW}), the saddle point of the action
$-kS^-_{WZW}(A_z)+\aksie \tr(A_z J_{\bar{z}})$ is at
$A_z^{(0)}(J_{\bar{z}})$, where $A_z^{(0)}$ is defined by the
equation $F(A_z^{(0)},\deel{1}{k}J_{\bar{z}})=0$\footnote{$F(A_z,
\abar)$ denotes the curvature of the connection
$\dif+A_z+\dbar+\abar$, and is given by $F=\dif\abar-\dbar A_z
+[A_z,\abar]$}. If we write $A_z=A_z^{(0)}+\tilde{A}_z$, and
$J_{\bar{z}}=-k\dbar h h^{-1}$, so that $A_z^{(0)}=-\dif h
h^{-1}$, then we can expand (\ref{def:sindWZW})
\be \label{def:sindWZW3}
\exp{-G[J_{\bar{z}}]}=
\int {\cal D} \tilde{A}_z \exp(-kS^-_{WZW}(h) - \aksies{k} \tr (
\tilde{A}_z \dif^{-1}_{A^{(0)}_z} \dbar_{\abar^{(0)}} \tilde{A}_z)
+\ldots),
\ee
where $\dif_{A_z^{(0)}}=\dif+\add(A_z^{(0)})$ and
$\dbar_{\abar^{(0)}}=\dbar+\add(\abar^{(0)})$, with
$\abar^{(0)}=\frac{1}{k}J_{\bar{z}}$.
This shows that the one-loop
contribution to $G[J_{\bar{z}}]$
is given by
$\hf\log\det(\dif^{-1}_{A_z^{(0)}}
\dbar_{\abar^{(0)}})$. If we {\it assume}
that this determinant is equal to
$\hf\log\det(\dbar_{\abar^{(0)}})-\hf\log\det(\dif_{A_z^{(0)}})$,
then we can
compute these determinants as explained below (\ref{eq:jac}),
to obtain
\ba \label{eq:sindone}
G_{one-loop}[J_{\bar{z}}] & = & - h_G S^+_{WZW}(\abar^{(0)}) +
h_G S^-_{WZW}(A_z^{(0)}) \nonu
& = & - h_G S^-_{WZW}(h) +  h_G S^+_{WZW}(h) \nonu
& = & -2 h_G S^-_{WZW}(h) + \aksies{h_G} \tr ( h^{-1}\dif h
h^{-1} \dbar h ).
\ea
The effective action up to one loop can be computed in the same
way as in (\ref{def:seffWZW})
\ba \label{def:seffWZW2}
S_{eff}(A_z) & = & \min_{h}
\left((k+2h_G)S^-_{WZW}(h^{-1}) - \aksies{h_G} \tr (h^{-1}\dif h
h^{-1} \dbar h) \right. \nonu & &  \;\;\;\;\;\; \left.
- \aksies{k} \tr (A_z   h^{-1}\dbar h)
\right) \nonu
& = & \min_{h'} \left( (k+2h_G)S^+_{WZW}(h') - \aksies{k+h_G}
\tr (A_z h'^{-1} \dbar h') \right),
\ea
where in the last line we changed variables from $h$ to $h'$,
with $h'^{-1}\dbar h'=(1-\deel{h_G}{k})h^{-1}\dbar h$. The
extremum is at $A_z=(1+\deel{h_G}{k})h'^{-1}\dif h'$, and we find
that up to one loop the effective action is given by
\be \label{eq:seff4}
S_{eff}(A_z)=-(k+2h_G) S^-_{WZW}((1-\deel{h_G}{k})A_z).
\ee

The disagreement between (\ref{eq:seff3}) and (\ref{eq:seff4}) is
quite puzzling. The one-loop calculation involves
the computation of a determinant, requiring a choice
of regularization procedure. For
2-d quantum gravity and $W_3$ gravity, a computation similar to
the one above agrees with independent
one-loop calculations performed in momentum space \cite{grini}.
In these momentum space calculations
one has to deal with momentum routing
ambiguities, and the agreement is only obtained after fixing
these in a rather ad hoc way. Nevertheless, this provides an
independent indication that (\ref{eq:seff4}) is a correct
result.
If one assumes that (\ref{eq:seff4}) is correct, then there is
something wrong with the derivation of (\ref{eq:seff3}). The
only non-classical step in this derivation is the replacement
${\cal D}A_z\rightarrow {\cal D}g$, and the only source of
trouble can be that the jacobian for the
change of variables $A_z\rightarrow g$ is not the same as the
jacobian for
the change of measures ${\cal D}A_z\rightarrow {\cal D}g$. Although
our understanding of how this could come about is incomplete,
the problem (if any)
seems to be related to giving a proper definition of
${\cal D}A_z$. The basic
property that fixes this measure is demanding that for arbitrary
functions $f(A_z)$,
\be \label{eq:basprop}
\int {\cal D}A_z \delta (A_z-X) f(A_z) = f(X).
\ee
However,
the
measure ${\cal D}g$ stems from the inner product
\be \label{def:gmes}
\langle\delta
g,\delta'g\rangle=\int d^2 z \, \sqrt{\det h_{ab}}\tr(g^{-1}\delta g
g^{-1}\delta'g),
\ee
where $h_{ab}$ is the two-dimensional metric.
The change of variables $g^{-1}\delta g \rightarrow \delta
A_z$ produces a measure ${\cal D}A_z$ coming from the
inner product
\be \label{def:ams}
\langle\delta A_z,\delta' A_z \rangle = \int d^2 z \,
\sqrt{\det h_{ab}} \tr(\delta A_z \delta' A_z),
\ee
And this inner product is ill-defined, because the integrand is not
a density, but something of conformal weight
$(\Delta,\bar{\Delta})=(3,1)$. A much more natural measure would
for instance be the one coming from the inner product
\be \label{def:ams2}
\langle\delta A_z,\delta' A_z \rangle = \int d^2 z \,
 \tr(\delta A_z \delta' A_{\bar{z}}(A_z)+
\delta' A_z \delta A_{\bar{z}}(A_z)),
\ee
where $A_{\bar{z}}(A_z)$ is such that the connection $A$ has
vanishing curvature. This measure is not consistent with
(\ref{eq:basprop}) and not with the
one-loop calculations, because the measure used there is determined
by treating $A_z$ as a free field, \ie\ one decomposes $A_z$ in
Fourier modes and defines the inner product such that these are
orthogonal. How one should compute the jacobian for the change
of measure from ${\cal D}g$ to such a measure is not clear.

To proceed,
we will assume that the effective action for the WZW
model acquires only multiplicative renormalizations, keeping the
above mentioned subtleties in mind.
If we do not assume this then the computation of
the effective action for $W_3$ gravity stops at (\ref{eq:seff2}).
In any case both
(\ref{eq:seff3}) and (\ref{eq:seff4}) are in agreement with this
assumption.
Thus, the rest of the computations are based on\\[5mm]
\noindent
{\bf conjecture 1}\\[2mm]
\noindent
The effective action for the WZW model is given by $-kZ_k
S^-_{WZW}(Z_A A_z)$, where $Z_k=1+{\cal O}(h_G/k)$ and
$Z_A=1+{\cal O}(h_G/k)$.\\[6mm]
As one can easily check with a calculation similar to the one
that led to (\ref{eq:seff3}), this conjecture follows if one assumes
that the following identity between path integrals is valid,
relating ${\cal D}A_z$ and ${\cal D}g$:\\[5mm]
\noindent
{\bf conjecture 1'}\\[2mm]
\noindent
For an arbitrary local functional $f$, \\
\noindent
$\int{\cal D}A_z\,\, f(A_z) \exp(-kS^-_{WZW}(A_z))=
 \int{\cal D} g \,\, f(Z_A^{-1} g^{-1}\dif g) \exp(-kZ_k
 S^-_{WZW}(g)).$ \\[6mm]
It is possible to prove the converse as well: conjecture 1
implies conjecture 1'. For the proof of this fact one first
decomposes
the function $f$ into Fourier modes, and then parametrizes an
arbitrary mode with a group valued variable $h$
via
\be f_h(A_z)=\exp(-\accc{Z_A Z_k k \dbar h h^{-1} A_z}).
\ee
Some
manipulations, using the Polyakov-Wiegmann identity and the
definition of the effective action, are then sufficient to
derive conjecture 1' for an arbitrary Fourier mode, and thus for
arbitrary functions $f$.

As far as the values of $Z_k$ and $Z_A$ are concerned, both the
calculation using the naive jacobian and the one-loop
calculation seem to suggest that $Z_k=1+\deel{2h_G}{k}+{\cal
O}(h_G/k)^2$. For
compact groups, the level of the WZW action must be an integer for
the action to be well defined, suggesting that the level does
not renormalize beyond one-loop, and that
$Z_k=(1+\deel{2h_G}{k})$ is indeed the full answer.
This is the value of $Z_k$ that we will use in the rest of the
paper. The same value for $Z_k$ was proposed in
\cite{pol4,wreview3}. As (\ref{eq:seff3}) and (\ref{eq:seff4}) predict
different values for $Z_A$ we will for $Z_A$ take some arbitrary
function $Z_A(k)$.



\newsection{The Effective Action of $W_3$ Gravity, Continued}

Using conjecture 1'  and $Z_k=(1+\frac{2h_G}{k})$
it is straightforward to work out the
effective action for $W_3$ gravity. Starting with
(\ref{eq:seff2}), and using that $h_G=3$ for $SL(3,\re)$, one
finds:
\ba
e^{-\Gamma_{eff}[T,W]} & = &
\int {\cal D} g \delta(J_1-\xi) \delta(J_2-\xi)
\delta(J_3) \delta(H_0) \delta(H_1) \delta (K_1-K_2) \nonu & &
\delta(T-N_T\xi(K_1+K_2))
\delta(W-N_W\xi^2K_3) e^{-kS_{WZW} (g)}
\nonu & = &
\int {\cal D} A_z \delta(J'_1-\xi) \delta(J'_2-\xi)
\delta(J'_3) \delta(H'_0) \delta(H'_1) \delta (K'_1-K'_2) \nonu & &
\delta(T-N_T\xi(K'_1+K'_2))
\delta(W-N_W\xi^2K'_3) e^{-(k-6)S_{WZW} (A_z)},
\label{eq:sefffin}
\ea
where $J'=kZ_A(k-6)A_z$. We can substitute the delta functions into
the WZW action, and obtain the effective action for $W_3$
gravity to all orders. The final result reads, in terms of the
renormalized level $k_c=k-6$:
\be \label{eq:w3full}
\Gamma_{eff}[T,W]=k_c S^-_{WZW}
\mat{0}{\frac{\xi}{
(k_c+6)Z_A(k_c)}}{0}{\frac{T}{2N_T \xi (k_c+6)Z_A(k_c)}}{0
}{\frac{\xi}{(k_c+6)Z_A(k_c)}}{\frac{W}{N_W \xi^2 (k_c+6)Z_A(k_c)}}{
\frac{T}{2N_T (k_c+6)Z_A(k_c)}}{0} .
\ee
The induced action for classical $W_3$ gravity, $\Gamma_L[T,W]$,
can also be obtained from a constrained WZW model; one can take
the large $k_c$ limit of (\ref{eq:w3full}), but one can also
directly compute the Ward identities for a constrained WZW model
and compare those with the Ward identities for the classical
$W_3$ algebra. In any case the result for $\Gamma_L[T,W]$ reads
\cite{kj1}
\be \label{eq:saddle}
\Gamma_{L}[T,W]=k S^-_{WZW}
\mat{0}{\alpha}{0}{\beta T}{0}{\alpha}{\gamma W}{\beta T}{0},
\ee
where $c=24k$, $2\alpha\beta k=-1$, and
$\gamma^2=-10\beta^2/\alpha^2$. Both (\ref{eq:w3full}) and
(\ref{eq:saddle}) contain one free parameter, and we can choose
$\xi/(k_c+6)Z_A(k_c)=\alpha=1$. This proves that
\be \label{eq:done}
\Gamma_{eff}[T,W]=Z_k \Gamma_L[Z_T T, Z_W W],
\ee
and using (\ref{def:c}) and (\ref{def:norm}) we find that $k_c$
and the central charge $c$ are related through
\be \label{def:c2}
c=50+24\left((k_c+3)+\frac{1}{(k_c+3)} \right)
\ee
and that the renormalizations $Z_k$, $Z_T$ and $Z_W$ are given
by
\ba \label{renorm}
Z_k & = & \frac{24}{c} k_c = 1-\frac{122}{c} +\ldots, \nonu
Z_T & = & \frac{c(k_c+3)}{24(k_c+6)^2Z_A(k_c)^2} , \nonu
Z_W & = & \frac{c\sqrt{(5c+22)}(k_c+3)^{3/2}}{
48\sqrt{30}(k_c+6)^{3}Z_A(k_c)^3}.
 \ea
These results are in agreement with the one-loop
results obtained in \cite{grini,vannie}, if
$Z_A(k_c)=1-\frac{3}{k_c}+{\cal O}(1/k_c)^2$, as predicted by
(\ref{eq:seff4}). Note that the `KPZ' relation between the level
$k_c$ and $c$ given in (\ref{def:c2}) is independent of $Z_A$,
and always comes out of this analysis as long as
$Z_k=1+\frac{2h_G}{k}$.
Clearly, the techniques used
here can be applied to $W_N$ gravity for arbitrary $N$, and in
particular to 2-d quantum gravity, yielding
\ba \label{grav}
\Gamma_{eff}[T] & = & Z_k \Gamma_L[Z_T T],\nonu
c & = & 13+6\left((k_c+2)+\frac{1}{(k_c+2)}\right) ,\nonu
Z_k & = & \frac{6}{c} k_c , \nonu
Z_T & = & \frac{c(k_c+2)}{6(k_c+4)^2Z_A(k_c)^2} .
\ea
These results agree with those obtained in \cite{kpz,zam,meipav,grini},
if $Z_A(k_c)=1-\frac{2}{k_c}+{\cal O}(1/k_c)^2$.

\newsection{Conclusions}

We have shown how one can obtain the effective action for $W_3$
gravity if the effective action for the WZW theory is known. It
is still an open problem to give a proof of conjecture 1,
or to show that it is false\footnote{It may also turn
out that the computation of the effective action of the WZW
model is so ambiguous, that one can {\it impose} conjecture 1 as
a regularization prescription}. A clue towards the
validity of conjecture 1 can be obtained by performing a
two-loop calculation for the WZW theory, which is under current
investigation \cite{privcom}.

Assuming the validity of conjecture 1, what could be
the exact values of $Z_k$ and $Z_A$? As we explained at the end
of section~4, it is reasonable to expect that
$Z_k=1+\frac{2h_G}{k}$, from which one can derive
the `KPZ' relation (\ref{def:c2})
between the central charge and the
renormalized level $k_c$ for $W_3$ gravity.
As for $Z_A$,
one might `argue' as follows (cf. \cite{wreview3}):
the effective action of the WZW
model has a current algebra of level $-k-2h_G$, and the fact
that the classical equation $(g{\cal J})=(-k-2h_G)\dif g$
renormalizes on the quantum level to $(g{\cal J})=(-k-h_G)\dif
g$ suggests that
\be \label{pos1}
Z_A(k)=\frac{k+h_G}{k+2h_G}=1-\frac{h_G}{k}+\ldots.
\ee
This value of $Z_A(k)$ has certain nice features: if we
substitute it in (\ref{renorm}) and (\ref{grav}), the
expressions given there simplify considerably and agree with
one-loop calculations. On the other hand, if conjecture 1' were
also valid for nonlocal functionals $f$, one could take
$f(A_z)=g(A_z)\exp (-lS^-_{WZW}(A_z))$ with an arbitrary
functional $g$, and evaluate the left hand side of conjecture 1'
in two different ways. It turns out that the two answers agree
for generic $g$ only if $Z_A=1$ and $Z_k=1+\frac{a}{k}$ for some
constant $a$. Thus, this suggests
\be
\label{pos2}
Z_A(k)=1,
\ee
leading to renormalization constants for gravity and $W_3$
gravity that disagree with the one-loop calculations. However,
these one-loop calculations were performed along the same lines
as the computation of (\ref{eq:seff4}), and as soon as one
claims that there is something wrong with (\ref{eq:seff4}), the
one-loop results for gravity become doubtful as well. Clearly,
a more precise analysis is needed to settle these issues.

The relation between the constrained WZW model presented here
and Toda theory becomes clear if one picks in
(\ref{def:seff}) the gauge choice $K_1=K_2=K_3=0$. Ignoring the
non-trivial contribution of the Faddeev-Popov ghosts in this
case, the action (\ref{def:seff}) reduces to a Toda action, and
$T$ and $W$ can be identified with the conserved currents of the
Toda theory.

All the computations in this paper have been done on the complex
plane. Working on a non-trivial Riemann surface will probably
introduce many extra subtleties, in particular one has to work
with generalized WZW actions \cite{wreview4}. We leave this and
other issues to future study.

\noindent {\bf Acknowledgement}\\[7mm]
\noindent We would like to thank B. de Wit for carefully
reading the manuscript. This work was financially supported
by the Stichting voor Fundamenteel Onderzoek der Materie (FOM).


\newpage

\begin{thebibliography}{99}


\bibitem{jj} J. de Boer and J. Goeree, \plb{274} (1992) 289.
\bibitem{jj2} J. de Boer and J. Goeree, \np{381} (1992) 329.
\bibitem{wreview1} C. M. Hull, 'Classical and Quantum
$W$-Gravity', QMW/PH/92/1.
\bibitem{wreview2} C. N. Pope, 'Lectures on $W$ Algebras and $W$
Gravity', CTP TAMU-103/91, Lectures given at the Trieste Summer
School in High-Energy Physics, August 1991.
\bibitem{wreview3} K. Schoutens, A. Sevrin and P. van Nieuwenhuizen,
'Induced Gauge Theories and $W$ gravity', ITP-SB-91-54, CERN-TH.6330/91,
LBL-31381, UCB-PTH-91/51, to appear in 'Strings and Symmetries
1991', Stony Brook, May 1991.
\bibitem{wreview4} J. de Boer and J. Goeree, 'The
Covariant Action and Its Moduli Space from Gauge Theory',
THU-92/14.
\bibitem{pol2} A. M. Polyakov, Mod. Phys. Lett. {\bf A2} (1987), 893.
\bibitem{meipav} K.A. Meissner, J. Pawelczyk, Mod. Phys. Lett.
{\bf A10} (1990) 763.
\bibitem{zam} Al. B. Zamolodchikov, preprint ITEP 84-89 (1989).
\bibitem{kpz} V. G. Knizhnik, A. M. Polyakov and A. B. Zamolodchikov,
Mod. Phys. Lett. {\bf A3} (1988) 819.
\bibitem{grini} M.T. Grisaru, P. van Nieuwenhuizen, `Loop
Calculations in Two-Dimensional Nonlocal Field Theories',
CERN-TH.6388/92.
\bibitem{beroog} M. Bershadsky and H. Ooguri, \cmp{126} (1989) 49.
\bibitem{sb-w3ind} K. Schoutens, A. Sevrin and P. van
Nieuwenhuizen, \np{364} (1991) 584.
\bibitem{drisok} V. G. Drinfel'd and V. V. Sokolov, J. Sov.
Math. {\bf 30} (1985) 1975, Sov. Math. Doklady {\bf 3} (1981)
457.
\bibitem{vannie} K. Schoutens, A. Sevrin and P. van
Nieuwenhuizen, \np{371} (1992) 315.
\bibitem{matsuo} Y. Matsuo, \plb{227} (1989) 209.
\bibitem{jj3} J. de Boer and J. Goeree, in preparation.
\bibitem{powie} A. M. Polyakov and P. B. Wiegmann, \plb{131}
(1983) 121; \plb{141} (1984) 233.
\bibitem{dublinann} J. Balog, L. Feh\'er, L. O'Raifeartaigh, P.
Forg\'acs and A. Wipf, Ann. Phys. {\bf 203} (1990) 76.
\bibitem{batvi} I.A. Batalin, G.A. Vilkovisky, \plb{102} (1981)
27; Phys. Rev. {\bf D28} (1983) 2567.
\bibitem{feifre} B. Feigin and  E. Frenkel \plb{246} (1990) 75;
E. Frenkel, '$W$-Algebras and Langlands-Drinfel'd
Correspondence', lectures given at Carg\`ese Summer School on
'New Symmetry Principles in QFT', July 1991.
\bibitem{fms} D. Friedan, E. Martinec and S. Shenker, \np{271}
(1986) 93.
\bibitem{kz} V. Knizhnik and A. B. Zamolodchikov, \np{247}
(1984) 83.
\bibitem{sander} F. Bais, P. Bouwknegt, M. Surridge and
K. Schoutens, \np{304} (1988) 348.
\bibitem{pol4} A. M. Polyakov, in Les Houches 1988, Fields,
Strings and Critical Phenomena, eds. E. Br\'ezin and J.
Zinn-Justin, North-Holland, 1990.
\bibitem{kj1} H. Ooguri, K. Schoutens, A. Sevrin and P. van
Nieuwenhuizen, `The Induced Action of $W_3$ Gravity',
ITP-SB-91/16, RIMS-764 (June 1991).
\bibitem{privcom} B. de Wit, M. Grisaru, P. van Nieuwenhuizen,
work in progress.

\end{thebibliography}
\end{document}